\documentclass[twocolumn,preprintnumbers]{revtex4}
\usepackage{amsmath}
\usepackage{graphicx}
\usepackage{dcolumn}
\usepackage{bm}
\usepackage{epsfig}

\begin{document}

\title{Six-dimensional Davidson potential as a dynamical symmetry \\
of the symplectic Interacting Vector Boson Model}
\author{H. G. Ganev$^{1}$, A. I. Georgieva$^{1,2}$, and J. P. Draayer$^{2}$}
\affiliation{$^1$Institute for Nuclear Research and Nuclear Energy, Bulgarian Academy of
Science, Sofia 1784, Bulgaria}
\affiliation{$^2$ Department of Physics and Astronomy, Louisiana State University, Baton
Rouge, LA 70803}

\begin{abstract}
A six-dimensional Davidson potential, introduced within the framework of the
Interacting Vector Boson Model (IVBM), is used to describe nuclei that
exhibit transitional spectra between the purely rotational and vibrational
limits of the theory. The results are shown to relate to a new dynamical
symmetry that starts with the $Sp(12,R) \supset SU(1,1) \times SO(6)$
reduction. Exact solutions for the eigenstates of the model Hamiltonian in
the basis defined by a convenient subgroup chain of $SO(6)$ are obtained. A
comparison of the theoretical results with experimental data for heavy
nuclei with transitional spectra illustrates the applicability of the theory.

\end{abstract}

\maketitle

\section{Introduction}

The interaction between competing collective modes in atomic nuclei is very
important in determining their structure. The collective modes that manifest
themselves most strongly \cite{exp} are rotations and vibrations. These
modes are characterized by very specific energy level spacings and
electromagnetic transition strengths. There are various models that give
exact algebraic solutions in these limits, one being the interacting boson
model (IBM) \cite{IBA} that contains both as special symmetry limits of the
overarching theory. Nonetheless, systems that exhibit a strongly mixed
rotational-vibrational character are neither easy to model nor understand,
even within an IBM-type algebraic picture that claims exact analytic results
in each of these symmetry limits.

The desire to have an algebraically solvable theory that can describe
systems with rotational-vibrational interactions has led nuclear physicist
to consider the Davidson potential \cite{Davidson}, which has known
algebraic solutions when applied to diatomic molecules. In an algebraic
approach for either the nuclear many-body problem or the Bohr-Mottelson
collective model, the addition of the Davidson potential to the Hamiltonian
requires the consideration of a dynamical subgroup chain that contains the
direct product $Sp(2,R) \otimes SO(n)\subset Sp(2n,R),$ with $n=3$ and $5$,
respectively. A short overview of these cases and their application in
nuclear physics is given in Section 2. In Section 3 the more general case of
a 6-dimensional Davidson potential is considered within the framework of the
phenomenological interacting vector boson model (IVBM) \cite{IVBM}. The
latter has $Sp(12,R)$ as its dynamical symmetry group. This model has been
applied successfully to a description of various collective phenomena \cite%
{IVBMrl,GGG,newprc}. In the present paper, a new reduction chain of the
dynamical group $Sp(12,R)$ through the direct product subgroup $%
Sp(2,R)\otimes SO(6)$ is reported. As shown in Section 4, this innovation
extends the applicability of the theory to include rotational-vibrational
interactions.

In short, the present study leads to a better understanding of and provides
motivation for an algebraic IVBM for two interacting many-particle systems.
The two systems of interest for nuclei are comprised of protons and
neutrons. The application of this dynamical symmetry to nuclei confirms the
ability of the Davidson potential to reproduce nuclear
rotational-vibrational behavior found in nature.

\section{Algebraic approaches in applications of the Davidson potential to
nuclear structure}

\subsection{The Davidson potential}

The need for a description of nuclei in which rotational-vibrational
interactions dominate has led to a search for algebraically solvable
potentials and a meaningful set of basis states that make the transitional
nature of these systems more transparent.

Davidson proposed such a potential,

\begin{equation}
V(r)=\chi (r^{2}+\frac{\varepsilon }{r^{2}}),  \label{DPdam}
\end{equation}
for diatomic molecules \cite{Davidson}. The Hamiltonian (including its
kinetic part) for a system with a strong rotational-vibrational interaction
in harmonic oscillator units takes the following form 
\begin{equation}
H_{\varepsilon }=\frac{1}{2}\hbar \omega ^{2}(-\nabla ^{2}+r^{2}+\frac{
\varepsilon }{r^{2}}).  \label{Hdam}
\end{equation}%
Both $\nabla ^{2}$ and ~$r^{2}$ are $SO(3)$\ scalars, and hence $H$ itself
is a $SO(3)$ invariant. On the other hand, $H$ can be expressed in terms of
the $SU(1,1)$ generators defined in the following way \cite{Rowe 2} 
\begin{equation*}
Z_{1}=-\nabla ^{2}+\frac{\varepsilon }{r^{2}},\quad Z_{2}=r^{2},\quad Z_{3}=%
\frac{1}{2}(r\cdot \nabla +\nabla \cdot r).
\end{equation*}
This means the eigenstates of the system can be classified according to the
direct product $SU(1,1) \times SO(3)$. Using the latter, algebraic solutions
(eigenvalues and wavefunctions) can be obtained. Indeed, as shown in \cite%
{Rowe 2}, the many-body system with the Davidson potential has $%
SU(1,1)\otimes SO(3)$ as its spectrum generating algebra.

It is well known in nuclear physics that the most successful description of
rotations and vibrations are obtained within the framework of the
Bohr-Mottelson collective model (CM) \cite{CM}, which in many respect is the
geometrical equivalent of the IBM \cite{IBA}. It is a liquid-drop model with
5 collective coordinates $(\nu =0,\pm 1,\pm 2)$ 
\begin{equation}
q_{\nu }=\beta \cos \gamma \mathcal{D}_{0\nu }^{2}(\Omega )+\frac{1}{\sqrt{2}
}\beta \sin \gamma (\mathcal{D}_{2\nu }^{2}(\Omega )+\mathcal{D}_{-2\nu
}^{2}(\Omega )\},  \label{CMcoor}
\end{equation}
and canonical momenta $\{p_{\nu }=-i\hbar \partial /\partial q_{\nu }$\},
expressed in terms of intrinsic ($\beta ,\gamma )$ coordinates and
rotational angles $\Omega $ of $SO(3)$. The model Hamiltonian 
\begin{equation*}
H_{0}=\frac{1}{2B}p\cdot p+\frac{1}{2}B\omega ^{2}q.q,
\end{equation*}%
gives a harmonic vibrational spectrum and with the addition of the
quadrupole Davidson potential 
\begin{equation}
H_{\varepsilon }=\frac{1}{2B}p\cdot p+\frac{1}{2}B\omega ^{2}\left( q.q+ 
\frac{\varepsilon }{q.q}\right),  \label{DPq}
\end{equation}%
yields a rotational-vibrational spectrum, characteristic of the so-called $%
\gamma-$soft Wilets-Jean rotor \cite{WJpot}. This limit of the model
corresponds to the $O(6)$-symmetry limit of the IBM \cite{IBA}. The
potentials used in these limits are independent of $\gamma $ (and $\Omega )$
and so are functions only of the variable $\beta ^{2}=q.q$, where both $p^{2}
$ and $\beta ^{2}$ are scalar products of 5-dimmensional vectors, so they
are $SO(5)$ invariants. Another nice feature of $H_{\varepsilon }$ (\ref{DPq}%
), is that it has $SU(1,1)$ dynamical symmetry. The energy spectrum for a $5$%
-dimensional collective Hamiltonian with a Davidson potential was given by
Elliott \textit{et al }\cite{Elliott}.

An orthonormal basis for the CM with the Davidson interaction is given \cite%
{RoHe} by a set of states $\{\left\vert nv\alpha LM\right\rangle\}$ that is
labelled by the quantum numbers of the subgroup chain

\begin{equation}
\begin{tabular}{ccccccc}
$SU(1,1)$ & $\otimes ~SO(5)$ & $\supset $ & $U(1)~\otimes $ & $SO(3)$ & $%
\supset $ & $SO(2)$ \\ 
$~n$ & $v$ & $\alpha $ & $n$ & $L$ &  & $M$%
\end{tabular}
,  \label{cho5}
\end{equation}
where $n$ labels the lowest weight $SU(1,1)$ state, $v$ is the highest
weight $SO(5)$ irreducible representation (irrep) label, and $\alpha$ is a
running index that is used to distinguish multiple occurrences of an $SO(3)$
irrep that is labelled by the angular momentum $L$ and its projection $M$.

Just as there is a correspondence between the description of the geometrical
collective model and the IBM for the $\gamma$-unstable transitional region,
an exact solution has also been obtained for the $O(6) \leftrightarrow U(5)$
transitional region using an infinite-dimensional algebraic technique \cite%
{FPJPD}. In \cite{FPJPD}, the complementarity of the $U(5)\supset SO(5)$ and 
$SO(6)\supset SO(5)$ bases to the $SU^{d}(1,1)\supset U(1)$ and $%
SU^{sd}(1,1)\supset U(1)$, respectively, with corresponding relations
between the Casimir invariants and quantum numbers labeling representations
in the corresponding chains, was used to obtain analytical results for the
energy eigenvalues and eigenstates of isotopic chains for transitional
nuclei lying between the $\gamma$-soft and vibrational limits of this theory.

\subsection{Collective behavior of many-body systems and symplectic geometry}

Notwithstanding these considerations, a microscopic description of the
collective behavior of a many-particle system in 3-dimensional space
requires higher symmetries. Specifically, such a complicated system is
usually characterized by an irrep of $Sp(6n,R)$ and its $Sp(6,R)\otimes O(n)$
subgroup, where $n=A-1$ and $A$ is equal to the total number of nucleons in
the system. Collective effects emerge within this structure when the system
is constrained to a specific irrep of $O(n)$ which in turn determines the $%
Sp(6,R)$ irrep \cite{CollGeom1}. When this is done, the Hamiltonian falls
within the enveloping algebra of $Sp(6,R)$ rather than $Sp(6n,R)$.

The coordinates $x_{is}$ and momenta $p_{is}$ of the $n$ particle system $%
s=1,2....,n$ in a 3-dimensional space $i=1,2,3$ defined by the only nonzero
commutator 
\begin{equation}
\quad \lbrack x_{is},p_{jt}]=i\delta _{ij}\delta _{st},\quad
i,j=1,2,3;~s,t=1,2,..,n\quad  \label{commut}
\end{equation}
are the elements of a $3n$-dimensional Weyl Lie algebra $W(3n)$. The
Hermitian quadratic expressions in the coordinates and momenta 
\begin{equation}
x_{is}x_{jt},\quad x_{is}p_{jt}+p_{jt}x_{is},\quad p_{is}p_{jt},
\label{Generators}
\end{equation}
that close under commutation \cite{Mosh1} yield the $3n(2\times 3n+1)$
generators of the $Sp(6n,R)$ symplectic group.

The generators of the subgroups in the reduction $Sp(6,R)\times O(n)$ of $%
Sp(6n,R)$ can be obtained from (\ref{Generators}) by means of contractions
with respect to the indices $s$ and $i$, respectively. For the $O(n)$ group
the infinitesimal operators have the \cite{Mosh2} well-known antisymmetrized
form 
\begin{equation}
\quad L_{st}=\sum_{i=1}^{3}(x_{is}p_{it}-x_{it}p_{is}).  \label{O(2)Gen}
\end{equation}
For $Sp(6,R)$ there are $3(2\times 3+1)=21$ Hermitian generators \cite{Mosh1}%
,\cite{AR}, \cite{RoweRos} of the form 
\begin{equation}
q_{ij}=\sum_{s=1}^{n}(x_{is}x_{js}),  \label{Sp6(a)}
\end{equation}%
\begin{equation}
S_{ij}=\frac{1}{2}%
\sum_{s=1}^{n}(x_{is}p_{js}+x_{js}p_{is}+p_{js}x_{is}+p_{is}x_{js}),
\label{Sp6(b)}
\end{equation}%
\begin{equation}
T_{ij}=\sum_{s=1}^{n}(p_{is}p_{js}),  \label{Sp6(c)}
\end{equation}
\begin{equation}
L_{ij}=\sum_{s=1}^{n}(x_{is}p_{js}-x_{js}p_{is}).  \label{Sp6(d)}
\end{equation}
The $L_{ij}$ generate the $SO(3)$ subgroup of $Sp(6,R)$ (\ref{Sp6(d)}). As
we are dealing only with the irreps $[\frac{1}{2}^{6}]$ or $[\frac{1}{2} ^{5}%
\frac{3}{2}]$ of $Sp(6,R)$, the irreps of $Sp(6,R)$ and $O(n)$\ are
complementary \cite{MoQu1}, i.e., if we fix the irrep of $O(n)$ that of $%
Sp(6,R)$ is specified or vice versa. From shell-model considerations it has
been shown \cite{vanagas} that fixing the $O(n)$ irrep isolates the
collective effects \cite{Dzublik}. Thus $Sp(6,R)$ can be used to solve the
collective part of the many-body problem.

A complete set of states is provided by the eigenstates of the Hamiltonian 
\begin{equation}
H_{0}=\sum_{s=1}^{n}\sum_{i=1}^{3}H_{is},\quad H_{is}=\frac{1}{2}
(x_{is}^{2}+p_{is}^{2}),  \label{H0}
\end{equation}
where the operators $H_{is}$ are special combinations of the generators of $%
Sp(6,R)$ that commute among themselves, i.e. $[H_{is},H_{jt}]=0$ and
therefore are the weight generators \cite{Mosh2} of this group. Furthermore,
in this basis we can calculate the matrix elements of all the generators of
the $Sp(6,R)$ group. This means that an arbitrary Hamiltonian $H$ involving
central forces is in the enveloping algebra of $Sp(6,R)$ and can be written
as a function of the quadratic expressions in (\ref{Generators}) which are
invariant under space reflections. This also applies to other integrals of
motion such as the square of components of the total angular momentum (\ref%
{Sp6(d)}) and functions thereof. Subgroups of the dynamical group $Sp(6,R)$
can be used to further classify the eigenstates of $H_{0}$.

The challenge is to define a basis in terms of irrep labels of groups in a
physically meaningful chain of subgroups of $Sp(6,R).$ The basis
characterized by the chain $Sp(6,R)\supset U(3)\supset SO(3),$ where $U(3)$
is the group of the quadrupole momentum introduced by Elliott, is well known
and obtained \ in both an abstract way \cite{RoweRos,MCC} as well as in
terms of shell-model states \cite{JPDSp6}. Many successful applications of
this theory have been made for deformed nuclear systems.

Another relevant chain for developing collective basis states is $%
Sp(6,R)\supset Sp(2,R)\otimes SO(3)$ that was considered by Moshinsky and
his collaborators in an effort to obtain a simple description of vibrational
collective nuclear motion \cite{CollGeom1}. Indeed if the local isomorphism
of the $sp(2,R)\approx su(1,1)$ algebras is taken into account its relation
to the spectrum generating algebra of the many body nuclear system with the
Davidson interaction becomes explicit. This provides motivation for
considering this reduction in seeking a description of a more complex modes
that includes rotational-vibrational interactions.

\section{New dynamical symmetry in the IVBM}

\subsection{Group theoretical background of the model}

On the basis of the above considerations, namely the use of a symplectic
geometry in investigating nuclear collective motion, a further elaboration
of the problem can be achieved if we consider the nuclear many-body system
as consisting of interacting proton and neutron subsystems. This leads to
the phenomenological IVBM \cite{IVBM}, where $Sp(12,R)$ -- the group of
linear canonical transformation in a $12$-dimensional phase space \cite%
{MoQu1} -- appears as the dynamical symmetry of the model. The $sp(12,R)$
algebra is realized in terms of creation (annihilation) operators $%
u_{m}^{+}(\alpha )(u_{m}(\alpha )),$ in a $3$-dimensional oscillator
potential $m=0,\pm 1$ of $two$ types of bosons differing by the value of the
``pseudo-spin'' projection $\alpha =p=1/2$ and $\alpha =n=-1/2$ . These are
related with the cyclic coordinates $x_{\pm 1}(\alpha )=\mp \frac{1}{\sqrt{2}%
}(x_{1}(\alpha )\pm ix_{2}(\alpha )),x_{0}(\alpha )=x_{3}(\alpha )$ and
their associated momenta $q_{m}(\alpha )=-i\partial /\partial x^{m}(\alpha )$
in the standard way 
\begin{eqnarray}
u_{m}^{+}(\alpha ) &=&\frac{1}{\sqrt{2}}(x_{m}(\alpha )-iq_{m}(\alpha )),
\label{cro} \\
u_{m}(\alpha ) &=&(u_{m}^{+}(\alpha ))^{\dagger }),  \notag
\end{eqnarray}
where $x_{i} (\alpha )\ i=1,2,3$ like in (\ref{commut}) are the Cartesian
coordinates of a quasi-particle vectors with an additional index, namely the
projection of the ``pseudo-spin'' $\alpha =\pm \frac{1}{2}$. The bilinear
products of the creation and annihilation operators of the two vector bosons
(\ref{cro}) generate the boson representations of the non-compact symplectic
group $Sp(12,R)$ \cite{IVBM}: 
\begin{eqnarray}
F_{M}^{L}(\alpha ,\beta ) &=&\underset{k,m}{\sum }C_{1k1m}^{LM}u_{k}^{+}(%
\alpha )u_{m}^{+}(\beta ),  \notag \\
G_{M}^{L}(\alpha ,\beta ) &=&\underset{k,m}{\sum }C_{1k1m}^{LM}u_{k}(\alpha
)u_{m}(\beta ),  \label{pairgen}
\end{eqnarray}
\begin{equation}
A_{M}^{L}(\alpha ,\beta )=\underset{k,m}{\sum }C_{1k1m}^{LM}u_{k}^{+}(\alpha
)u_{m}(\beta ),  \label{numgen}
\end{equation}
where $C_{1k1m}^{LM}$ are the usual Clebsch-Gordon coefficients and $L=0,1,2$
and $M=-L,-L+1,...L$ define the transformation properties of (\ref{pairgen})
and (\ref{numgen}) under rotations. The commutation relations between the
pair creation and annihilation operators \ (\ref{pairgen}) and the number
preserving operators (\ref{numgen}) are calculated in \cite{IVBM}. The set
of operators (\ref{numgen}) close under commutation to form the algebra of
the maximal compact subgroup of $Sp(12,R)\supset U(6).$ The linear invariant
of $U(6)$ is the number operator, 
\begin{equation}
N=\sqrt{3}\bigskip (A^{0}(p,p)+A^{0}(n,n))=N_{+}+N_{-},  \label{Nt}
\end{equation}
that counts the total number of bosons. The action space of the operators (%
\ref{pairgen}) and (\ref{numgen}) is in general reducible and the invariant
operator $(-1)^{N}$ decomposes it into even \textrm{H}$_{+}$ with $%
N=0,2,4,...,$ and odd \textrm{H}$_{-}$ with $N=1,3,5,...,$ subspaces, so in
the reduction $Sp(12,R)\supset $ $U(6)$ both the even and odd irreps of the $%
Sp(12,R)$ decompose into an infinite sum of finite fully symmetric irreps of 
$U(6)$, $[N]_{6}=[N,0^{5}]_{6}$ , where $N$ is the eigenvalue of the
operator (\ref{Nt}).

\subsection{Reduction through the non-compact Sp(2,R)}

In order to relate the IVBM to the $6-$dimensional Davidson potential, we
introduce another reduction of the $Sp(12,R)$ group through its non-compact
subgroup \cite{CollGeom1}, \cite{MoQu1}, \cite{vanagas}: 
\begin{equation}
Sp(12,R)\supset Sp(2,R)\otimes SO(6).\   \label{nred}
\end{equation}
As can be deduced from the considerations given above, this construction
obviously survives the addition of Davidson potential. The infinitesimal
generators of the $Sp(2,R)$ algebra 
\begin{equation*}
F=\underset{k,m,\alpha }{\sum }C_{1k1m}^{00}u_{k}^{+}(\alpha
)u_{m}^{+}(\alpha )=2S^{+},
\end{equation*}
\begin{equation}
G=\underset{k,m,\alpha }{\sum }C_{1k1m}^{00}u_{k}(\alpha )u_{m}(\alpha
)=2S^{-},  \label{Sp2Gen}
\end{equation}
\begin{equation*}
A=\underset{k,m,\alpha }{\sum }C_{1k1m}^{00}u_{k}^{+}(\alpha )u_{m}(\alpha )=%
\frac{1}{\sqrt{3}}N=2S^{0}-1,
\end{equation*}
are obtained from the $Sp(12,R)$ generators (\ref{pairgen}) and (\ref{numgen}%
) by means of contraction with respect to both the spatial $m=0,\pm 1$ and \
the \textquotedblright pseudospin\textquotedblright\ $\alpha =p=1/2$, $%
\alpha =n=-1/2$ indices. It is straightforward to show that the operators $%
S^{\tau },\tau =0,\pm $ commute in a standard way for the $SU(1,1)$ algebra
generators \cite{FPJPD} 
\begin{equation*}
\lbrack S^{0},S^{\pm }]=S^{\pm },\;[S^{+},S^{-}]=-2S^{0},
\end{equation*}%
so the $sp(2,R)$ and the $su(1,1)$ algebras are locally isomorphic with a
Casimir operator written as $C_{2}(SU(1,1))=S^{0}(S^{0}-1)-S^{+}S^{-}.$

By construction, the operators (\ref{Sp2Gen}) are scalars with respect to
6-dimensional rotations and they commute with the components of the
6-dimensional momentum operators \cite{IVBM}, 
\begin{equation}
\Lambda _{M}^{L}(\alpha ,\beta )=A_{M}^{L}(\alpha ,\beta
)-(-1)^{L}A_{M}^{L}(\beta ,\alpha ),  \label{O6Gen}
\end{equation}
which obey the property %\bigskip
\begin{equation*}
\Lambda _{M}^{L}(\alpha ,\beta )=(-1)^{L}\Lambda _{M}^{L}(\beta ,\alpha )
\end{equation*}%
and generate the $SO(6)$ algebra. When $\alpha =\beta $, from (\ref{O6Gen})
one obtains 
\begin{equation*}
\Lambda _{M}^{L}(\alpha ,\alpha )=A_{M}^{L}(\alpha ,\alpha
)-(-1)^{L}A_{M}^{L}(\alpha ,\alpha ),
\end{equation*}
which are different from zero only for $L=1$. Indeed, the $6$ operators $%
A_{M}^{1}(p,p)$ and $A_{M}^{1}(n,n)$ are rank one tensors with respect to $%
O(3)$ and represent, respectively, the angular momenta of the $p$ and $n$
boson systems. In the case when $\alpha \neq \beta $ the operators (\ref%
{O6Gen}) are the $5+3+1$ components of tensors of rank $2$, $1$, and $0$,
respectively. In this way, the direct product of the two groups (\ref{nred})
is realized. The second order invariant for the $SO(6)$\ group is 
\begin{equation}
\Lambda ^{2}=\underset{L,\alpha ,\beta }{\sum }(-1)^{M}\Lambda
_{M}^{L}(\alpha ,\beta )\Lambda _{-M}^{L}(\beta ,\alpha ),  \label{C2O6}
\end{equation}%
and it is related to the second order invariant of the $Sp(2,R),$ as in the
direct product (\ref{nred}) the two groups are complementary \cite{MoQu1},
which means that the irreps of the group $SO(6)$ determine those of $Sp(2,R)
\approx SU(1,1)$ and vice versa.

\subsection{Labeling of the basis}

In order to define the basis of the system with (\ref{nred}) as a dynamical
symmetry that allows one to include the $6$-dimensional Davidson potential,
we consider the reduction of the $SO(6)$ algebra to the $SO(3)$ algebra of
the angular momentum through the following chain \cite{IVBM}, \cite{ChCF} 
\begin{equation}
\begin{tabular}{lllllll}
$SO(6)$ & $\supset $ & $SU(3)$ & $\otimes $ & $O(2)$ & $\supset $ & $SO(3)$
\\ 
$\omega $ &  & $(\lambda ,\mu )$ &  & $\nu $ &  & $L$%
\end{tabular}%
,  \label{redO6}
\end{equation}
which defines the $\gamma$-unstable limit of the IVBM. The single
infinitesimal operator of $O(2)$ is proportional to the scalar operator $%
\Lambda ^{0}(\alpha ,\beta )$ from the $SO(6)$ generators (\ref{O6Gen}), 
\begin{equation}
M_{\alpha \beta }=-\sqrt{3}\Lambda ^{0}(\alpha ,\beta )=-\sqrt{3}%
[A^{0}(\alpha ,\beta )-A^{0}(\beta ,\alpha )],  \label{o2gen}
\end{equation}
and the generators of $SU(3)$ \cite{IVBM} are 
\begin{equation*}
X_{M}^{2}=i(A_{M}^{2}(p,n)-A_{M}^{2}(n,p)),M=0,\pm 1,\pm 2,
\end{equation*}%
\begin{equation}
Y_{M}^{1}=A_{M}^{1}(p,p)+A_{M}^{1}(n,n)=-\frac{1}{\sqrt{2}}L_{M},M=0,\pm 1.
\label{su3gen}
\end{equation}
Note, that in this case the quadrupole moment $X$ (\ref{su3gen}) is the
proton-neutron interaction. The second-order Casimir invariants of the two
groups in the direct product in (\ref{redO6}) can be written as 
\begin{equation*}
2C_{2}(O_{2})=M^{2}=\underset{\alpha ,\beta }{\sum }M_{\alpha \beta
}M_{\beta \alpha },
\end{equation*}
\begin{equation*}
C_{2}(SU(3))=\underset{M}{\sum }(-1)^{M}(X_{M}X_{-M}+Y_{M}Y_{-M}).
\end{equation*}

For $SO(6)\subset U(6)$, the symmetric representation $[N]_{6}$ of $U(6)$
decomposes into fully symmetric $(\omega ,0,0)_{6}\equiv (\omega )_{6}$
irreps of $SO(6)$ \cite{VanBook} according to the rule 
\begin{equation}
\lbrack N]_{6}=\bigoplus_{\omega =N,N-2,...,0(1)}(\omega
,0,0)_{6}=\bigoplus_{i=0}^{<\frac{N}{2}>}(N-2i)_{6},  \label{U6O6}
\end{equation}
where $<\frac{N}{2}>=\frac{N}{2}$ if $N$ is even and $\frac{N-1}{2}$ if $N$
is odd. Furthermore, the following relation between the quadratic \ Casimir
operators $C_{2}(SU(3))$, $M^{2}$ of $\ O(2)$\ and $\Lambda ^{2}$ (\ref{C2O6}%
) of $SO(6) $ holds \cite{Dragt}: 
\begin{equation}
\Lambda ^{2}=2C_{2}(SU(3))-\frac{1}{3}M^{2},  \label{Casimirs}
\end{equation}%
which means that the reduction from $SO(6)$ to the rotational group $SO(3)$
is carried out through the complementary groups $O(2)$\ and $SU(3)$ \cite%
{MoQu1}. As a consequence, the irrep labels $[f_{1},f_{2},0]_{3}$ of $SU(3)$
are determined by $(\omega )_{6}$ of $SO(6)$\ and by the integer label $(\nu
)_{2}$ of the associated irrep of $O(2)$ i.e. 
\begin{equation}
(\omega )_{6}=\bigoplus [f_{1},f_{2},0]_{3}\otimes (\nu )_{2}.  \label{o6su3}
\end{equation}%
Using the relation (\ref{Casimirs}) of the Casimir operators, for their
respective eigenvalues one obtains: 
\begin{equation}
\omega (\omega +4)=\frac{4}{3}(f_{1}^{2}+f_{2}^{2}-f_{1}f_{2}+3f_{1})-\frac{%
\nu ^{2}}{3}.  \label{wfv}
\end{equation}
Then if $f_{2}=0$ and $\nu =f_{1}$, from (\ref{wfv}) written as $%
(f_{1}-\omega )(f_{1}+\omega +4)=0$, it follows that $f_{1}=\omega $. If $%
f_{2}=i$ in (\ref{wfv}) we obtain the relation \ 
\begin{equation}
\nu =\pm (\omega -2i),\quad i=0,1,...,\omega .  \label{O6O2}
\end{equation}%
%\
Hence (\ref{o6su3}) can be rewritten as 
\begin{eqnarray*}
(\omega )_{6} &=&\bigoplus_{i=0}^{\omega }[\omega ,i,0]_{3}\otimes (\omega
-2i)_{2}= \\
&&\bigoplus_{\nu =\omega ,\omega -2,...,0(1)}[\omega ,\frac{\omega -\nu }{2}%
,0]_{3}\otimes (\nu )_{2},
\end{eqnarray*}%
or in terms of the Elliott's notation $(\lambda ,\mu )$ 
\begin{equation}
(\omega )_{6}=\bigoplus_{\nu =\omega ,\omega -2,...,0(1)}(\frac{\omega +\nu 
}{2},\frac{\omega -\nu }{2})\otimes (\nu )_{2}.  \label{O6U3}
\end{equation}

Finally, the convenience of this reduction can be further enhanced through
the use of the standard rules for the reduction of the $SU(3)\supset SO(3)$
chain:

\begin{eqnarray}
K &=&\min (\lambda ,\mu ),\min (\lambda ,\mu )-2,...,0~(1)  \notag \\
L &=&\max (\lambda ,\mu ),\max (\lambda ,\mu )-2,...,0~(1);K=0  \label{su3o3}
\\
L &=&K,K+1,...,K+\max (\lambda ,\mu );K\neq 0.  \notag
\end{eqnarray}
The latter is the usual reduction of an irrep $(\lambda,\mu)$ of $SU(3)$
into irreps $L$ of $SO(3)$ where the multiplicity number $K$ is used to
label the rotational bands in the energy spectra of the system.

\subsection{A 6-dimensional Davidson potential and its algebraic structure}

As discussed in Sec.2, basis states of the collective model are classified
according to irrep labels (quantum numbers) of the reductions \ \cite{Rowe 2}%
, \cite{Rowe DS}\ 
\begin{equation*}
\begin{tabular}{lllllllll}
$SU(1,1)$ & $\otimes $ & $SO(5)$ & $\supset $ & $U(1)$ & $\otimes $ & $SO(5)$
& $\supset $ & $SO(3).$%
\end{tabular}%
\end{equation*}
For spherical nuclei, $SU(1,1)$ is realized when $U(1)\otimes SO(5)$ is a
subgroup of $U(5)$; on the other hand, for deformed nuclei a transformation
of $SU(1,1)$ gives eigenstates of the Davidson potential. With the extension
of the collective model to six\ dimensions, the states can be classified by
the quantum numbers \cite{Rowe DS} provided by the chain 
\begin{equation*}
\begin{tabular}{lllllll}
$U(3)$ & $\supset $ & $U(1)$ & $\otimes $ & $SU(3)$ & $\supset $ & $SO(3)$%
\end{tabular}%
\end{equation*}%
which are eigenstates of the spherical vibrational Hamiltonian 
\begin{equation*}
H_{0}=h\omega N.
\end{equation*}%
The $U(1)\otimes SU(3)$ subalgebra, generating a rotational type of
spectrum, appears also in the chain 
\begin{equation*}
\begin{tabular}{lllllllll}
$SU(1,1)$ & $\otimes $ & $SU(3)$ & $\supset $ & $U(1)$ & $\otimes $ & $SU(3)$
& $\supset $ & $SO(3).$%
\end{tabular}%
\end{equation*}
It should be clear that one can make the same transformation of $SU(1,1)$ as
for the $5-$dimensional collective model, and in so doing obtain basis
states which are eigenstates of a $6-$dimensional analog of the Davidson
potential. The former reduction chain is contained naturally in the
group-theoretical structure of the Interacting Vector Boson Model \cite%
{CollGeom1}, \cite{MoQu1}, \cite{vanagas}: 
\begin{equation}
\begin{tabular}{lllllllll}
$Sp(12,R)$ & $\supset $ & $Sp(2,R)$ & $\otimes $ & $SO(6)$ &  &  &  &  \\ 
&  & $N$ &  & $\omega $ &  &  &  &  \\ 
&  & $\wr \wr $ &  & $\cup $ &  &  &  &  \\ 
&  & $SU(1,1)$ & $\otimes $ & $SU(3)$ & $\otimes $ & $O(2)$ & $\supset $ & $%
SO(3),$ \\ 
&  & $N$ &  & $(\lambda ,\mu )$ &  & $\nu $ & $K$ & $L$%
\end{tabular}
\label{correspondence}
\end{equation}
where below the different subgroups the quantum numbers which characterize
their irreducible representations are shown. In the last line of (\ref%
{correspondence}) we take into account the fact that the group $Sp(2,R)$ is
locally isomorphic to the group $SU(1,1)$. The basis, labeled by the quantum
numbers classified by the group-subgroup chain (\ref{correspondence}), can
be written as 
\begin{equation}
|N\omega ;(\lambda ,\mu )\nu ;K,L\rangle  \label{basis}
\end{equation}%
where the reduction rules for obtaining specific values for each state are
the same as given earlier. By means of these labels, the basis states can be
classified in each of the two irreducible even \textrm{H}$_{+}$ with $%
N=0,2,4,...,$ and odd \textrm{H}$_{-}$ with $N=1,3,5,...$ representations of 
$Sp(12,R)$. We illustrate this in Table 1 for the even \textrm{H}$_{+}$
irreducible representation, where $N$ with the set of $\omega $ contained in
it (\ref{U6O6}) label the rows and the values of the quantum number $\nu $ ( %
\ref{O6O2}) label the columns. The $SU(3)$ quantum numbers $(\lambda ,\mu )$
define the cells of the Table 1 as they are obtained with the help of $%
\omega $ and $\nu $ (\ref{O6U3}).

The Hamiltonian with the dynamical symmetry corresponding to the chain (\ref%
{correspondence}) is expressed in terms of the first and second order
Casimir operators of the different subgroups in it: \ 
\begin{equation}
H=aN+bN^{2}+\alpha _{6}\Lambda ^{2}+\alpha _{2}M^{2}+\beta _{3}L^{2}.
\label{Halmitonian}
\end{equation}%
and it is obviously diagonal in the basis (\ref{basis}) labeled by the
quantum numbers of their representations. The second order invariant of $%
SU(3)$ is dropped in (\ref{Halmitonian}), because of its linear dependence
on the Casimir operators of the $SO(6)$ and \ $O(2)$ (\ref{Casimirs}). Then
the eigenvalues of the Hamiltonian (\ref{Halmitonian}) that yield the
spectrum of a system interacting with $6-$dimensional Davidson potential
are: 
\begin{eqnarray}
E(N,\omega ,\nu ,L) &=&aN+bN^{2}+\alpha _{6}\omega (\omega +4)  \notag \\
&&+\alpha _{2}\nu ^{2}+\beta _{3}L(L+1).  \label{Energies}
\end{eqnarray}%
\begin{widetext}
\begin{center}
\textbf{Table 1.}
\begin{equation*}
\begin{tabular}{||c|c|ccccccccc}
\hline\hline
$N$ & $\omega $ & $\nu \cdots $ & $\quad 6$ & $4$ & $2$ & $0$ & $-2$ & $-4$
& $-6\quad $ & $\cdots $ \\ \hline
\multicolumn{1}{||l|}{$0$} & $0$ & \multicolumn{1}{|l}{} &
\multicolumn{1}{l}{} & \multicolumn{1}{l}{} & \multicolumn{1}{l}{} &
\multicolumn{1}{l}{$(0,0)$} & \multicolumn{1}{l}{} & \multicolumn{1}{l}{}
& \multicolumn{1}{l}{} & \multicolumn{1}{l}{} \\ \cline{1-2}\cline{6-8}
\multicolumn{1}{||l|}{$2$} &
\begin{tabular}{l}
$2$ \\
$0$
\end{tabular}
& \multicolumn{1}{|l}{} & \multicolumn{1}{l}{} & \multicolumn{1}{l}{} &
\multicolumn{1}{l}{
\begin{tabular}{l}
$(2,0)$ \\
\end{tabular}
} & \multicolumn{1}{l}{$
\begin{tabular}{l}
$(1,1)$ \\
$(0,0)$%
\end{tabular}
\ $} & \multicolumn{1}{l}{%
\begin{tabular}{l}
$(0,2)$ \\
\end{tabular}
} & \multicolumn{1}{l}{} & \multicolumn{1}{l}{} & \multicolumn{1}{l}{} \\
\cline{1-2}\cline{5-9}
\multicolumn{1}{||l|}{$4$} &
\begin{tabular}{l}
$4$ \\
$2$ \\
$0$%
\end{tabular}
& \multicolumn{1}{|l}{} & \multicolumn{1}{l}{} & \multicolumn{1}{l}{
\begin{tabular}{l}
$(4,0)$ \\
\\
\end{tabular}
} & \multicolumn{1}{l}{
\begin{tabular}{l}
$(3,1)$ \\
$(2,0)$ \\
\end{tabular}
} & \multicolumn{1}{l}{
\begin{tabular}{l}
$(2,2)$ \\
$(1,1)$ \\
$(0,0)$
\end{tabular}
} & \multicolumn{1}{l}{
\begin{tabular}{l}
$(1,3)$ \\
$(0,2)$ \\
\end{tabular}
} & \multicolumn{1}{l}{
\begin{tabular}{l}
$(0,4)$ \\
\\
\end{tabular}%
} & \multicolumn{1}{l}{} & \multicolumn{1}{l}{} \\ \cline{1-2}\cline{4-10}
\multicolumn{1}{||l|}{$6$} &
\begin{tabular}{l}
$6$ \\
$4$ \\
$2$ \\
$0$
\end{tabular}
& \multicolumn{1}{|l}{} & \multicolumn{1}{l}{
\begin{tabular}{l}
$(6,0)$ \\
\\
\\
\end{tabular}
} & \multicolumn{1}{l}{
\begin{tabular}{l}
$(5,1)$ \\
$(4,0)$ \\
\\
\end{tabular}
} & \multicolumn{1}{l}{
\begin{tabular}{l}
$(4,2)$ \\
$(3,1)$ \\
$(2,0)$ \\
\end{tabular}
} & \multicolumn{1}{l}{
\begin{tabular}{l}
$(3,3)$ \\
$(2,2)$ \\
$(1,1)$ \\
$(0,0)$%
\end{tabular}
} & \multicolumn{1}{l}{
\begin{tabular}{l}
$(2,4)$ \\
$(1,3)$ \\
$(0,2)$ \\
\end{tabular}
} & \multicolumn{1}{l}{
\begin{tabular}{l}
$(1,5)$ \\
$(0,4)$ \\
\\
\end{tabular}
} & \multicolumn{1}{l}{
\begin{tabular}{l}
$(0,6)$ \\
\\
\\
\end{tabular}
} & \multicolumn{1}{l}{} \\ \hline
\multicolumn{1}{||l|}{$\vdots $} &  & \multicolumn{1}{|l}{$\vdots $} &
\multicolumn{1}{l}{$\vdots $} & \multicolumn{1}{l}{$\vdots $} &
\multicolumn{1}{l}{$\vdots $} & \multicolumn{1}{l}{$\vdots $} &
\multicolumn{1}{l}{$\vdots $} & \multicolumn{1}{l}{$\vdots $} &
\multicolumn{1}{l}{$\vdots $} & \multicolumn{1}{l}{$\ddots $} \\ \hline\hline
\end{tabular}
\end{equation*}
\end{center}
\end{widetext}

\section{Application to real nuclei}

In applications of this new dynamical symmetry (\ref{correspondence}) of the
IVBM to real nuclear systems that follows, we exploit the ``algebraic''
definition of yrast states as introduced in \cite{GGG}. This means that we
consider those states with maximal value of the angular momentum $L$ for a
given number of bosons $N$ to be yrast. With this definition, the states of
the ground band, which are the yrast states of the nucleus, are basis states
with $\omega =N$, $\lambda =\mu = \frac{N}{2}$ where $N=0,4,8,...$ ($\Delta
N=4$) from the $\nu =0$ column of Table 1.

The states of excited bands are not yrast states. The correct placement of
the bands in the spectrum strongly depends on their band-heads'
configuration, and in particular, on the number of bosons, $N=N_{\min }$,
from which they are built \cite{newprc}. For the excited bands considered,
we choose first a corresponding $N_{\min }$ for the band-head state and then
the bands are developed by changing $N_{\min }$ with $\Delta N=4$ and $%
\Delta \omega =2,$ so that the lowest $L$ of the band-head is taken from the 
$N_{\min }$ multiplet and $\Delta L=2$ for the $\beta _{i}-$bands with $%
K_{i}^{\pi }=0_{i}^{+}$, where $i$ enumerates the $0^{+}$ excited states in
the order of increasing energy and $\Delta L=1$ for $K^{\pi }=L^{+}~(L\neq 0$%
, $L=2,4,..)$, as prescribed by the reduction rules (\ref{su3o3}). The
values of the $(\lambda,\mu )$ of the $su(3)$ multiplets to which the
excited bands correspond are obtained by fixing the value of the $o(2)$
label $\nu $. For example, the states of first excited $\beta-$band and/or $%
\gamma$-band may belong to two different diagonals $(\lambda ,\mu =0)$, $K=0$
and/or to $(\lambda,\mu =2)$, $K=0$ and/or $2,$ so that $\nu =L$ or/and $\nu
=L-2$ for $L$-even and $\nu =L-1$ for $L$-odd respectively and $\Delta \nu =2
$ for each neighboring $su(3)$ multiplets in a band under consideration.
This variety of possible choices for the excited bands allows us to
reproduce correctly the behavior of these bands with respect to one another,
which can change a lot even in neighboring nuclei because of the mixing of
the vibrational and rotational collective modes \cite{GP}.

From (\ref{Energies}) it is obvious that there are $5$ free parameters,
which we determine by fitting the theoretical predictions for the energies
of the ground and few excited bands to the experimental data \cite{exp} ,
using a $\chi^{2}-$procedure. The parameters that were obtained, the number $%
s$ of experimental states, $\chi ^{2}$, and $N_{\min }$ are all given in
Table 2 for four different nuclei.

\begin{center}
\textbf{Table 2.} 
\begin{equation*}
\begin{tabular}{||l||l|l|l|l|l||}
\hline\hline
Nucleus & $s$ & $N_{\min }$ & bands & $\chi ^{2}$ & parameters \\ 
\hline\hline
$^{168}$Yb & $20$ & $0$ & ground & $0.0047$ & 
\begin{tabular}{l}
$a=0.03493$ \\ 
$b=0.00059$%
\end{tabular}
\\ 
$R_{4}=3.26$ & $6$ & $20$ & $\gamma $ & $0.0006$ & $\alpha _{6}=0.00055$ \\ 
& $4$ & $20$ & $\beta _{1}$ & $0.0073$ & $\alpha _{2}=-0.00245$ \\ 
& $3$ & $24$ & $\beta _{2}$ & $0.0125$ & $\beta _{3}=0.00229$ \\ \hline
$^{232}$Th & $15$ & $0$ & ground & $0.0050$ & 
\begin{tabular}{l}
$a=0.01986$ \\ 
$b=0.00037$%
\end{tabular}
\\ 
$R_{4}=3.28$ & $9$ & $24$ & $\gamma $ & $0.0002$ & $\alpha _{6}=0.00035$ \\ 
& $4$ & $20$ & $\beta _{1}$ & $0.0022$ & $\alpha _{2}=-0.00096$ \\ 
& $2$ & $28$ & $\beta _{2}$ & $0.0054$ & $\beta _{3}=0.00145$ \\ 
& $3$ & $38$ & $K=4$ & $0.0020$ &  \\ \hline
$^{150}$Sm & $11$ & $0$ & ground & $0.0057$ & 
\begin{tabular}{l}
$a=0.09705$ \\ 
$b=0.00016$%
\end{tabular}
\\ 
$R_{4}=2.32$ & $5$ & $12$ & $\gamma $ & $0.0089$ & $\alpha _{6}=0.00042$ \\ 
& $3$ & $6$ & $\beta _{1}$ & $0.0083$ & $\alpha _{2}=0.00064$ \\ 
& $3$ & $12$ & $\beta _{2}$ & $0.0198$ & $\beta _{3}=0.00123$ \\ \hline
$^{152}$Sm & $9$ & $0$ & ground & $0.0022$ & 
\begin{tabular}{l}
$a=0.02971$ \\ 
$b=0.00057$%
\end{tabular}
\\ 
$R_{4}=3.01$ & $7$ & $24$ & $\gamma $ & $0.0023$ & $\alpha _{6}=0.00056$ \\ 
& $7$ & $14$ & $\beta _{1}$ & $0.0105$ & $\alpha _{2}=0.00229$ \\ 
& $3$ & $28$ & $\beta _{2}$ & $0.0088$ & $\beta _{3}=0.00229$ \\ \hline\hline
\end{tabular}%
\end{equation*}
\end{center}

\begin{figure}[th]
\centerline{\hbox{\epsfig{figure=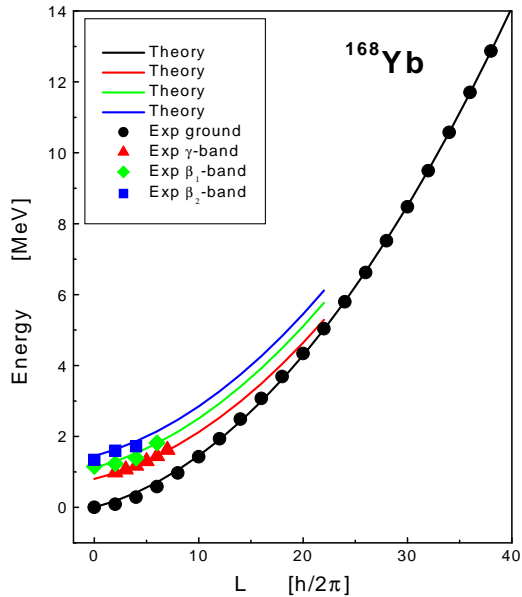,height=10cm}}}
\caption{(Color online) Comparison of the theoretical and experimental
energies for ground and excited bands of $^{168}$Yb.}
\label{Yb}
\end{figure}

\begin{figure}[th]
\centerline{\hbox{\epsfig{figure=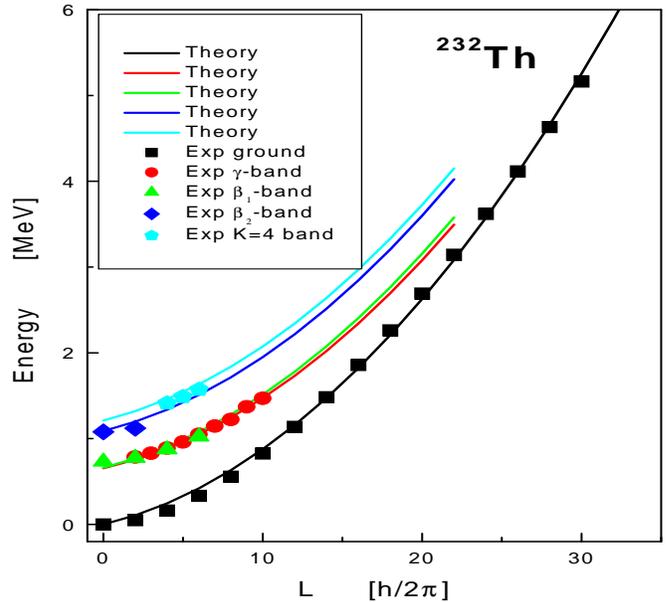,width=10cm,height=10cm}}}
\caption{(Color online) The same as in Fig. 1, but for the nucleus $^{232}$%
Th.}
\label{Th}
\end{figure}

\begin{figure}[th]
\centerline{\hbox{\epsfig{figure=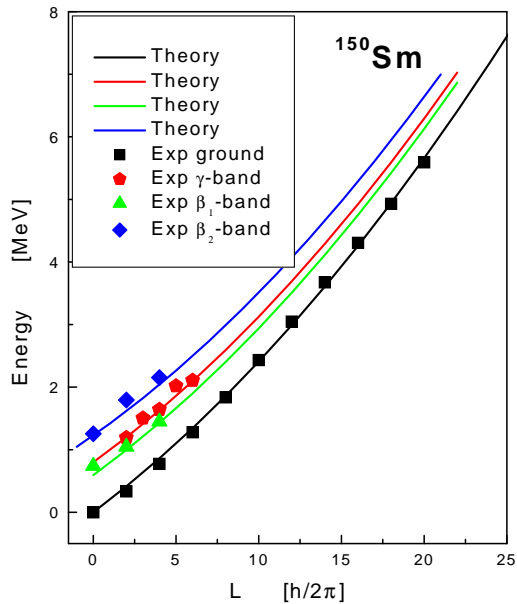,height=10cm}}}
\caption{(Color online) Comparison of the theoretical and experimental
energies for $^{150}$Sm.}
\label{Sm150}
\end{figure}

\begin{figure}[th]
\centerline{\hbox{\epsfig{figure=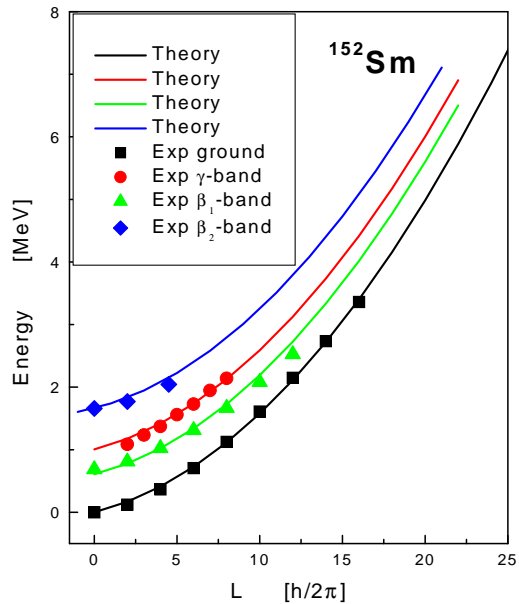,height=10cm}}}
\caption{(Color online) The same as on Fig. 1 for $^{152}$Sm.}
\label{Sm152}
\end{figure}

The first two, $^{168}$Yb and $^{232}$Th, are rather well deformed nuclei
from the rare earth and light actinide region. We have chosen these
particular nuclei as they have rather long spin sequences going up to very
high values of the angular momentum in their ground bands. From the
experimental ratios $R_{4}=\frac{E_{4}}{E_{2}}$ of the ground band energies,
which is $R_{4}=3.26$ for the $^{168}$Yb and $R_{4}=3.28$ for the $^{232}$%
Th, we see that these two nuclei are very close to rotators, especially in
their ground state bands. The values of $N_{\min }$ that determine the start
of the band-heads for the excited bands considered are rather high. One can
see the good agreement between theory and experiment for the ground, first
two $\beta$ and $\gamma$ bands for $^{168}$Yb, and the ground, first two $%
\beta$, $\gamma$ and $K=4$ bands for $^{232}$Th, respectively, in Figure \ref%
{Yb} and Figure \ref{Th}. The parameters (see Table 2) in both cases are of
the same order of magnitude, but for $^{168}$Yb they are somewhat larger. In
both cases the $0^{+}$ band-head of the $\beta$-band is placed below the $%
2^{+}$ band-head of the $\gamma$-band. Additionally, one can clearly see the
different degree of degeneracy of the $L$-even states of the $\gamma$ and
first $\beta$ bands, which is a typical property for well-deformed nuclei.

Additionally, in Table 2 we give the results for the $^{150}$Sm and $^{152}$%
Sm, which are considered to be transitional nuclei. The first one with $%
R_{4}=2.32$ is transitional between the $\gamma$-soft $(R_{4}=2.5)$ and
vibrational nuclei with $(R_{4}=2.0)$ and the second one is a nuclei at the
critical point of phase/shape transition \cite{X5exp} with so-called $X(5)$
symmetry. As shown in Figure \ref{Sm150} and \ref{Sm152}, the experimental
data is reproduced remarkably well, especially for $^{152}$Sm where the $%
\beta$ and $\gamma$ bands are well distinguished. The values of $N_{\min }$
varies more for these two nuclei and the parameter $\alpha _{2}$ changes its
sign.

The values of $\chi^{2}$ are rather good (small) for all of the examples
considered. This suggests that the model is appropriate for the description
of a rather broad range of nuclei, and most importantly nuclei that display
mixed rotational and vibrational degrees of freedom.

\section{Summary and \textbf{conclusions}}

In this work we introduce a new reduction of the dynamical group $Sp(12,R)$
of the algebraic Interacting Vector Boson Model. It is based on the fact,
that the rotational-vibrational spectrum of the nuclear system can be
generated from a Hamiltonian with a Davidson interaction which has as a
spectrum generating algebra the direct product $SU(1,1)\otimes SO(n)$. The
three- and five-dimensional cases ($n$ = $3$ and $5$, respectively) have
already been explored in the nuclear structure physics and are related to
the many-body problem in 3-dimensional space and the $\gamma$-soft limit of
the geometrical Collective Model, respectively. Based on the algebraic
approaches to these problems, we introduce an extension of the spectrum
generating algebra to $SU(1,1)\otimes SO(6)$ which includes the $6$%
-dimensional Davidson potential. It is naturally contained in the group of
dynamical symmetry $Sp(12,R)$ of the IVBM (\ref{correspondence}). Further,
the reduction of the boson representations of $SU(1,1)\otimes SO(6)$ to the
angular momentum group $SO(3)$ is obtained in order to provide for a
complete labeling of the basis states of the system and the model
Hamiltonian is written in terms of the first and second order invariants of
the groups from the corresponding reduction chain. Hence the problem is
exactly solvable within the framework of the IVBM which, in turn, yields a
simple and straightforward application to real nuclear systems.

We present results that were obtained through a phenomenological fit of the
models' predictions for the spectra of collective states to the experimental
data for two nuclei from the rare-earth and actinide major shells exhibiting
rotational spectra, as well as two with transitional character between $%
\gamma$-soft and vibrational spectra. The good agreement between the
theoretical predictions and the experiment results confirms the
applicability of the IVBM to a broad range of nuclei with quite different
collective properties. These features could be further developed to study
the phase/shape transitions in nuclear systems \cite{RoPRL}, which of late
has been a subject of high interest from a theoretical \cite{newsym} as well
as from experimental \cite{X5exp} point of view.

The most important feature of the model, from a physical point of view, is
that it leads to a successful description of different types of nuclear
collective spectra as well as mixed-mode results with the proton and neutron
substructures and interactions between them taken into account explicitly.
This is accomplished within the framework of a symplectic symmetry that
allows for a change in the number of bosons of each type.

\section*{Acknowledgments}

Discussions with D. J. Rowe, G. Rosensteel, D. Bonatsos and P. van Isaker
are gratefully acknowledged. This work was supported by the U.S. National
Science Foundation, Grant Number 0140300.

\end{document}